\documentclass[runningheads,a4paper]{llncs}

\usepackage{amssymb}
\usepackage{graphicx}
\usepackage[T1]{fontenc}
\usepackage{multirow}
\usepackage{amsmath}

\usepackage{url}
\urldef{\mailsa}\path|{placzek.bartlomiej, marcin.bernas}@gmail.com|    
\newcommand{\keywords}[1]{\par\addvspace\baselineskip
\noindent\keywordname\enspace\ignorespaces#1}

\begin{document}

\mainmatter  %

\title{Data suppression algorithms for surveillance applications of wireless sensor and actor networks}

\titlerunning{Data suppression algorithms for surveillance applications}
\author{Bart\l{}omiej P\l{}aczek \and Marcin Bernas}
\authorrunning{B. P\l{}aczek, M. Bernas}
\institute{University of Silesia, Institute of Computer Science,\\
B\k{e}dzi\'nska 39, 41-200 Sosnowiec, Poland\\
\mailsa\\}

\toctitle{Data suppression algorithms for surveillance applications of wireless sensor and actor networks}
\tocauthor{Bart\l{}omiej P\l{}aczek, Marcin Bernas}
\maketitle

\begin{abstract}
This paper introduces algorithms for surveillance applications of wireless sensor and actor networks (WSANs) that reduce communication cost by suppressing unnecessary data transfers. The objective of the considered WSAN system is to capture and eliminate distributed targets in the shortest possible time. Computational experiments were performed to evaluate effectiveness of the proposed algorithms. The experimental results show that a considerable reduction of the communication costs together with a performance improvement of the WSAN system can be obtained by using the communication algorithms that are based on spatiotemporal and decision aware suppression methods.\footnote{Preprint of: P\l{}aczek, B., Bernas, M.: Data Suppression Algorithms for Surveillance Applications of Wireless Sensor and Actor Networks. in Gaj, P., Kwiecien, A., Stera, P. (eds.) Computer Networks. CCIS, vol. 522, pp. 23-32, 2015. The final publication is available at link.springer.com} 
\end{abstract}

\section{Introduction}

Wireless sensor and actor networks (WSANs) are composed of sensor nodes and actors that are coordinated via wireless communications to perform distributed sensing and acting tasks. In WSANs, sensor nodes collect information about the physical world, while actors use the collected information to take decisions and perform appropriate actions upon the environment. The sensor nodes are usually small devices with limited energy resources, computation capabilities and short wireless communication range. In contrast, the actors are equipped with better processing capability, stronger transmission powers and longer battery life. The number of actors in WSAN is significantly lower than the number of sensor nodes \cite{bibbp1,bibbp2}.

The WSANs technology has enabled new surveillance applications, where sensor nodes detect targets of interest over a large area. The information collected by sensor nodes allows mobile actors to achieve surveillance goals such as target tracking and capture. Several examples of the WSAN-based surveillance applications can be found in the related literature, including land mine destruction \cite{bibbp3}, chasing of intruders \cite{bibbp4}, and forest fires extinguishing \cite{bibbp5}.

The surveillance applications of WSANs require real-time data delivery to provide effective actions. A fast response of actors to sensor inputs is necessary. Moreover, the collected information must be up to date at the time of acting. On the other hand, the sensor readings have to be transmitted to the mobile actors through multi-hop communication links, which results in transmission delays, failures and random arrival times of packets. The energy consumption, transmission delay, and probability of transmission failure can be reduced by decreasing the amount of transmitted data \cite{bibbps1,bibbps2,bibbpmb}. Thus, minimization of data transmission is an important research issue for the development of the WSAN-based surveillance applications \cite{bibbp2}. It should be noted that other methods can be used in parallel to alleviate the above issues, e.g., optimisation of digital circuits design for network nodes \cite{bibbppp}.

This paper introduces an approach to reduce the data transmission in WSAN by means of suppression methods that were originally intended for wireless sensor networks (WSNs). The basic idea behind data suppression methods is to send data to actors only when sensor readings are different from what both the sensor nodes and the actors expect. In the suppression schemes, a sensor node reports only those data readings that represent a deviation from the expected behaviour. Thus, the actor is able to recognize relevant events in the monitored environment and take appropriate actions.

The data suppression methods available in the literature were designed for monitoring applications of WSNs. In such applications, a sink node needs to collect information describing a given set of parameters with a defined precision or recognize predetermined events. These state-of-the-art suppression methods are based on an assumption that a large subset of sensor readings does not need to be reported to the sink as these readings can be inferred from the other transferred data \cite{bibbp6,bibbp7,bibbp8,bibbp9}. In order to infer suppressed data, the sink uses a predictive model of the monitored phenomena. The same model is used by sensor nodes to decide if particular data readings have to be transmitted. A sensor node suppresses transmission of a data reading only when it can be inferred within a given error bound.

Temporal suppression techniques exploit correlations between current and historical data readings of a single sensor node. The simplest scheme uses a na\"{\i}ve model, which assumes that current sensor reading is the same as the last reported reading \cite{bibbp10}. When using this method, a sensor node transmits its current reading to sink only if difference between the current reading and previously reported reading is above a predetermined threshold.

Parameters monitored by WSNs usually exhibit correlations in both time and space \cite{bibbp11}. Thus, several more sophisticated spatiotemporal suppression methods were proposed that combine the basic temporal suppression with detection of spatially correlated data from nearby nodes \cite{bibbp8,bibbp9,bibbp12}. According to the spatiotemporal approach, sensor nodes are clustered based on spatial correlations. Sensor readings within each cluster are collected at a designated node (cluster head), which then uses a spatiotemporal model to decide if the readings have to be transmitted to sink.

In previous work of the first author \cite{bibbp13} a decision-aware data suppression approach was proposed, which eliminates transfers of sensor readings that are not useful for making control decisions. This approach was motivated by an observation that for various control tasks large amounts of sensor readings often do not have to be transferred to the sink node as control decisions made with and without these data are the same. The decision-aware suppression was used for optimizing transmission of target coordinates from sensor nodes to a mobile sink which has to track and catch a moving target. According to that approach only selected data are transmitted that can be potentially useful for reducing the time in which the target will be reached by the sink.

According to the authors' knowledge, there is a lack of data suppression methods in the literature dedicated for the surveillance applications of WSANs. In this paper the available data suppression methods are adapted to meet the requirements of the WSANs. Effectiveness of these methods is evaluated by using a model of WSAN, where mobile actors have to capture randomly distributed targets in the shortest possible time. 

The paper is organized as follows. Details of the WSAN model are discussed in Section 2. Section 3 introduces algorithms that are used by actors to navigate toward targets as well as algorithms of sensor - actor communication that are based on the data suppression concept. Results of simulation experiments are presented in Section 4. Finally, conclusions are given in Section 5.

\section{Network model}

In this study a model of WSAN is considered, which includes 16 actors and 40000 sensor nodes. The monitored area is modelled as a grid of 200 x 200 square segments. Discrete coordinates $(x, y)$ are used to describe positions of segments, sensor nodes, actors, and targets ($x = 0, 1, ..., 199$, $y = 0, 1, ..., 199$). The sensor nodes are placed in centres of the segments. Each sensor node detects presence of a target in a single segment. Communication range of a sensor node covers the segment where this node is located as well as the eight neighbouring segments. Radius of the actor's communication range equals 37 segments. In most cases, the sensor nodes have to use multi-hop transmission for reporting their readings to actors. Due to the long communication range, each actor can transmit data to a large number of nodes (up to 4293) directly in one hop. 

The task of sensor nodes is to detect stationary targets in the monitored area and report their positions to actors. On the basis of the received information, each actor selects the nearest target and moves toward it. This process is executed in discrete time steps. Maximum speed of actor equals two segments per time step. At each time step three new targets are created at random positions. A target is eliminated if an actor reaches the segment in which the target was detected. The targets may correspond to fires, intruders, landmines, enemy units, etc.

\begin{figure}
\centering
\includegraphics [width=6cm] {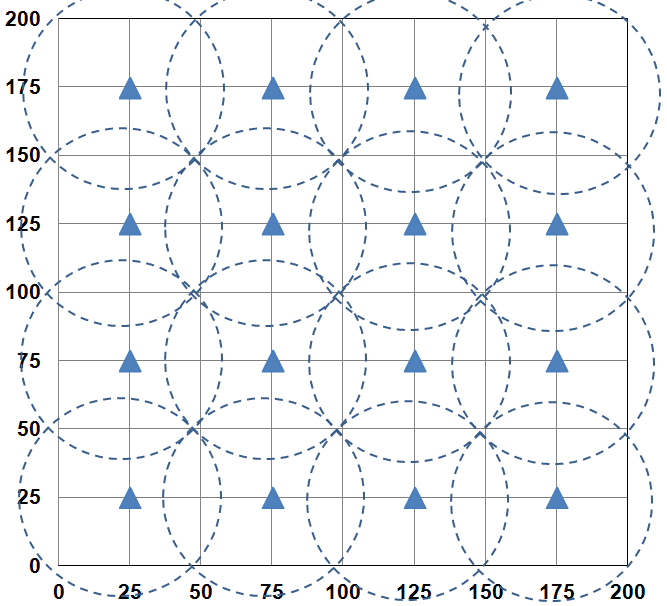}
\caption{Default positions (triangles) and communication ranges (circles) of actors}
\label{fig:fig1bp}
\end{figure}

Default (initial) positions of actors were determined to ensure that the communication ranges of the 16 actors cover the entire monitored area (Fig.~\ref{fig:fig1bp}). An actor, which has not received information from sensor nodes about current target locations moves toward its default position. Such situation occurs when there is no target within the actor's range or the information about detected target is suppressed by sensor node. 

For the above WSAN model, data communication cost is evaluated by using two metrics: number of data transfers (packets sent from sensor nodes to actors), and total hop count. The hop count is calculated assuming that the shortest path is used for each data transfer. Performance of the targets elimination by actors is assessed on the basis of average time to capture, i.e., the time from the moment when a target is created to the moment when it is captured by an actor and eliminated.  

\section{Actor navigation and data communication algorithms}

During target chasing, the mobile actors decide their movement directions based on the navigation algorithm, which is presented in Tab.~\ref{tab:tab1bp}. Each actor holds a target map $TM$ to collect the information delivered by particular sensor nodes. An element of target map $TM(x, y)_i$ equals 1 if the $i$-th actor has received information that there is target detected in segment $(x, y)$. In opposite situation, the target map element equals 0. The following symbols are used in the pseudo-code of the navigation algorithm: $(x_\mathrm{C}, y_\mathrm{C})_i$ denotes a segment in which the $i$-th actor is currently located, $(x_\mathrm{N}, y_\mathrm{N})_i$ is the nearest target according to the information collected by $i$-th actor in its target map, $(x_\mathrm{S}, y_\mathrm{S})_i$ is the selected destination segment, and $(x_\mathrm{D}, y_\mathrm{D})_i$ denotes the default actor position. It should be remembered that discrete coordinates $(x, y)$ are used to identify the segments. Current position of actor, the destination segment as well as the targets map are broadcasted by the actor to all sensor nodes in its communication range.

An actor moves toward its default position unless a target is registered in its target map. Each actor takes a decision regarding the segment $(x_\mathrm{C}^+, y_\mathrm{C}^+)_i$ into which it will move during the next time step. The actor's decision is taken by solving the following optimization problem:
\begin{align}
  \textnormal{minimze } d((x_\mathrm{C}^+, y_\mathrm{C}^+)_i,(x_\mathrm{S}, y_\mathrm{S})_i) \nonumber \\
  \textnormal{subject to } d((x_\mathrm{C}^+, y_\mathrm{C}^+)_i,(x_\mathrm{C}, y_\mathrm{C})_i)\leq v_{max}
\end{align}
where $d(\cdot)$ denotes the Euclidean distance between segments and $v_{max} = 2$ (segments per time step) is the maximum speed of actor.

\begin{table}
\centering
\caption{Pseudo-code of navigation algorithm executed by $i$-th actor}
\begin{tabular}{ll}
\hline
 1 & at each time step do \\ 
 2 & \hspace{0.4cm} $TM(x_\mathrm{C}, y_\mathrm{C})_i := 0$ \\ 
 3 & \hspace{0.4cm} broadcast $(x_\mathrm{C}, y_\mathrm{C})_i$, $(x_\mathrm{S}, y_\mathrm{S})_i$, and $TM_i$ \\ 
 4 & \hspace{0.4cm} collect data from sensor nodes and update $TM_i$ \\ 
 5 & \hspace{0.4cm} if at least one target is registered in $TM_i$ then \\ 
 6 & \hspace{0.8cm} find the nearest target $(x_\mathrm{N}, y_\mathrm{N})_i$ in $TM_i$ \\ 
 7 & \hspace{0.8cm} $(x_\mathrm{S}, y_\mathrm{S})_i := (x_\mathrm{N}, y_\mathrm{N})_i$ \\ 
 8 & \hspace{0.4cm} else  \\ 
 9 & \hspace{0.8cm} $(x_\mathrm{S}, y_\mathrm{S})_i := (x_\mathrm{D}, y_\mathrm{D})_i$ \\ 
 10 & \hspace{0.4cm} move toward $(x_\mathrm{S}, y_\mathrm{S})_i$ \\
\hline 
\end{tabular} 
\label{tab:tab1bp}
\end{table}

The sensor nodes report their readings to the actors by using the algorithm presented in Tab.~\ref{tab:tab2bp}. Each sensor node holds a list of actors $AL_{(x, y)}$ which includes IDs $(i)$ of actors that have communicated their status data to the sensor node during current time step. It means that the sensor placed in segment $(x, y)$ knows the actual positions, destinations and target maps of the actors listed in $AL_{(x, y)}$.

In order to minimize the data communication cost it was assumed that at each time step a sensor node may transmit data to one selected actor $(i^*)$. Two simple actor selection conditions are considered in this study. According to the first actor selection condition, a sensor node reports its reading to the nearest known actor. The rationale behind this condition is that the nearest actor is expected to capture the target in shortest time and the additional benefit is that the transmission requires minimum hop count. The second actor selection condition aims at balancing the actors' workload. When using this condition, a sensor node selects that actor for which the number of targets registered in the target map is minimal.

\begin{table}
\centering
\caption{Pseudo-code of data communication algorithm executed by sensor node $(x, y)$}
\begin{tabular}{ll}
\hline
 1 & at each time step do \\ 
 2 & \hspace{0.4cm} collect $(x_\mathrm{C}, y_\mathrm{C})_i$, $(x_\mathrm{S}, y_\mathrm{S})_i$, and $TM_i$ from actors and update $AL_{(x, y)}$ \\ 
 3 & \hspace{0.4cm} if target is detected and was not reported to any actor then \\ 
 4 & \hspace{0.8cm} find actor $i^*$ in $AL_{(x, y)}$ which satisfies actor selection condition \\ 
 5 & \hspace{0.8cm} if suppression condition is not satisfied then \\ 
 6 & \hspace{1.2cm} report target position to actor $i^*$ \\ 
 7 & \hspace{0.4cm} if target was reported to actor $i^*$ and is no longer detected then \\ 
 8 & \hspace{0.8cm} if actor $i^*$ is available in $AL_{(x, y)}$ and $TM(x, y)_{i^*} := 1$ then \\ 
 9 & \hspace{1.2cm} report elimination of target to actor $i^*$ \\ 
\hline 
\end{tabular} 
\label{tab:tab2bp}
\end{table}

The data communication algorithm presented in Tab.~\ref{tab:tab2bp} utilizes the temporal data suppression method to minimize the amount of transmitted data. The temporal suppression is introduced by the if-then statements in lines 3 and 7 of the pseudo-code (Tab.~\ref{tab:tab2bp}). This basic data suppression method is extended to spatiotemporal and decision-aware suppression by using additional condition (5-th line of the pseudo-code). 

According to the spatiotemporal data suppression method (STS), a sensor node in segment $(x, y)$ will suppress reporting the position of detected target to the selected actor $i^*$ if the information which is currently available for the actor indicates that there is at least one target within distance $d_\mathrm{STS}$ from segment $(x, y)$. The distance threshold $d_\mathrm{STS}$ is calculated according to the following formula: 
\begin{equation}
 d_\mathrm{STS} = \alpha \cdot d((x_\mathrm{C}, y_\mathrm{C})_{i^*},(x, y)), 
\end{equation}
where $\alpha$ is a parameter of the algorithm.

The above approach is based on the heuristic rule that the actor does not need the precise information about target location to chase the target effectively when the distance to target is large. The closer to target, the higher precision of the localization has to be obtained \cite{bibbp14}.

In case of the decision aware suppression method, the transmission of target coordinates $(x, y)$ from the sensor node to actor $i^*$ is suppressed if it can be expected that this information will not influence the actor's decision. It means that the sensor node suppresses data transmission if, according to the information available for the node, the actor will move into the same segment $(x_\mathrm{C}^+, y_\mathrm{C}^+)_i$ regardless of whether the information about the new target in segment $(x, y)$ is transmitted or not. For instance, the suppression is performed when there is a target located in the actor's destination segment $(x_\mathrm{S}, y_\mathrm{S})_{i^*}$ and the new detected target $(x, y)$ is more distant to the actor than segment $(x_\mathrm{S}, y_\mathrm{S})_{i^*}$.
 
Similarly like for the spatiotemporal suppression, the distance between sensor and actor is taken into account when using the decision aware suppression method. The suppression is performed only if this distance is above a predetermined threshold value $d_\mathrm{DAS}$ (in segments).

\section{Experiments}

Computational experiments were performed to compare the data communication cost and target chasing performance for six algorithms that use different combinations of the actor selection and data suppression approaches (Tab.~\ref{tab:tab3bp}). The comparison was made by taking into account three criteria: time to capture, hop count and number of data transfers. The WSAN model presented in Section 2 was used for a simulation-based evaluation of the above metrics. During simulations the targets were created at random time steps and locations. The simulation is finished when the number of eliminated targets reaches 2700. This section discusses the experimental results that were obtained from 20 simulation runs for each algorithm and parameter setting.

\begin{figure}
\centering
\includegraphics [width=7cm] {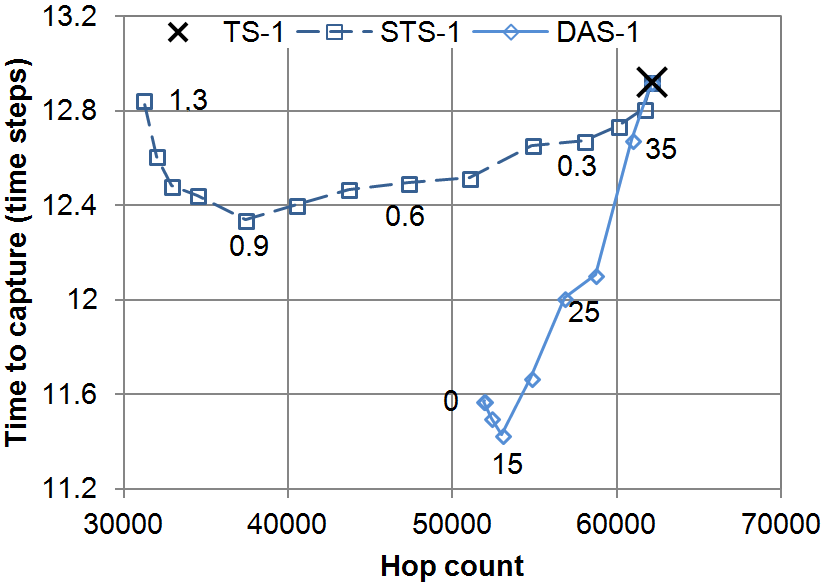}
\caption{Time to capture vs. hop count for algorithms TS-1, STS-1, and DAS-1}
\label{fig:fig2bp}
\end{figure}

\begin{figure}
\centering
\includegraphics [width=7cm] {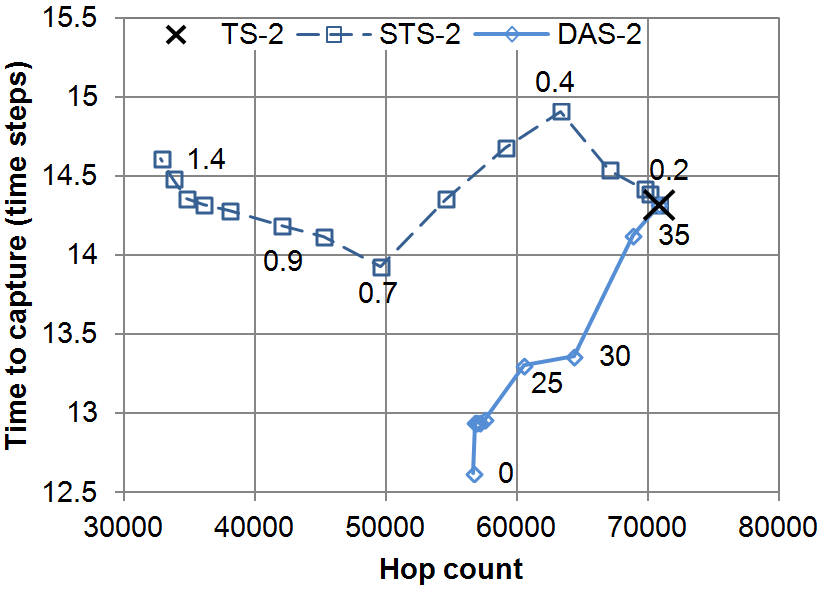}
\caption{Time to capture vs. hop count for algorithms TS-2, STS-2, and DAS-2}
\label{fig:fig3bp}
\end{figure}

Figures \ref{fig:fig2bp} and \ref{fig:fig3bp} depict the dependencies between average values of hop count and time to catch for the compared algorithms. Labels of data points in the charts show values of the algorithm parameters. The results of STS-1 and STS-2 algorithms are presented for $\alpha$ ranging between 0 and 1.4. In case of DAS-1 and DAS-2 the distance threshold $d_\mathrm{DAS}$ was changed from 0 to 40 in steps of 5 segments. In general, lower communication cost (hop count) and higher performance (shorter time to capture) were achieved by the algorithms from the first group (TS-1, STS-1, and DAS-1), in which the target is always reported to the nearest actor. It can be observed in Figs. \ref{fig:fig2bp} and \ref{fig:fig3bp} that the spatiotemporal and decision aware suppression methods improve the results of the temporal suppression. 

\begin{table}
\centering
\caption{Compared algorithms}
\begin{tabular}{p{2cm}ll}
\hline
 Algorithm & Actor selection condition & Data suppression method \\
\hline 
 TS-1 & \multirow{3}{*}{Nearest actor} & Temporal \\ 
 STS-1 & & Spatiotemporal \\ 
 DAS-1 & & Decision aware \\ 
\hline 
 TS-2 & \multirow{3}{*}{\parbox[t]{5cm}{Actor with minimum number \\ of reported targets}} & Temporal \\ 
 STS-2 & & Spatiotemporal \\ 
 DAS-2 & & Decision aware \\
\hline
\end{tabular} 
\label{tab:tab3bp}
\end{table}

It should be noted that for $\alpha = 0$ the STS algorithm corresponds to TS. Similarly, DAS corresponds to TS for $d_\mathrm{DAS} > 37$ as the radius of actor's communication range equals 37. The shortest time to capture was achieved by using the DAS-1 algorithm with $d_\mathrm{DAS} = 15$ segments. The lowest hop counts were obtained for the STS algorithms with high $\alpha$ values.

\begin{figure}
\centering
\includegraphics [width=11cm] {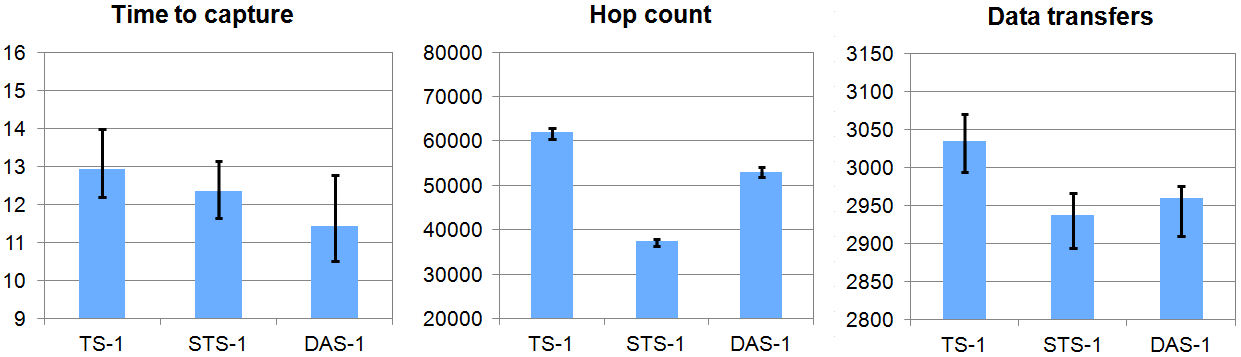}
\caption{Time to capture, hop count and number of data transfers for algorithms TS-1, 
STS-1 ($\alpha = 0.9$) , and DAS-1 ($d_{DAS} = 15$)}
\label{fig:fig4bp}
\end{figure}

\begin{figure}
\centering
\includegraphics [width=11cm] {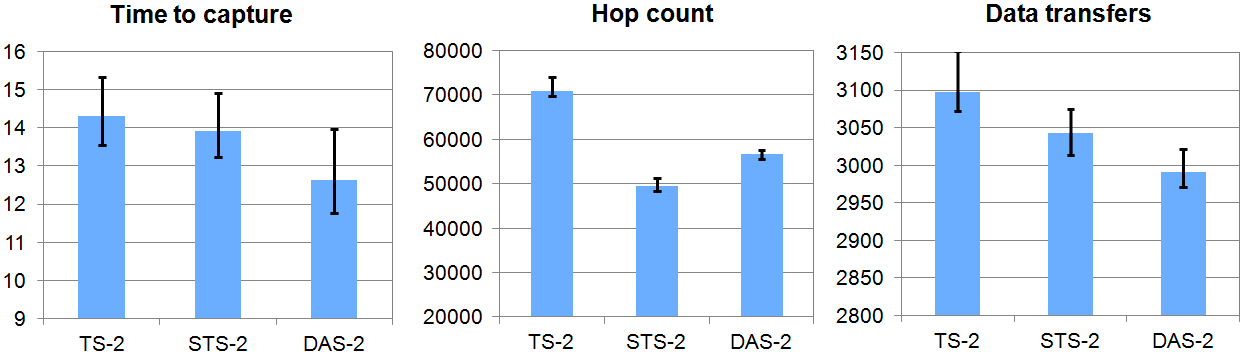}
\caption{Time to capture, hop count and number of data transfers for algorithms TS-2, STS-2 ($\alpha = 0.7$), and DAS-2 ($d_{DAS} = 0$)}
\label{fig:fig5bp}
\end{figure}

Figures \ref{fig:fig4bp} and \ref{fig:fig5bp} present detailed results for selected settings that allow the compared algorithms to achieve the maximum performance, i.e., minimum average time to capture. The error bars show the range between minimum and maximum of the metrics obtained from the 20 simulation runs. Average values are depicted as columns.

According to the presented results, it can be concluded that the spatiotemporal and decision aware suppression methods reduce the number of data transfers and hop counts in comparison with the temporal suppression. Moreover, these approaches decrease the average time in which actors eliminate the targets. The effect of decreased time to capture is especially visible for the algorithms that are based on the decision aware suppression. The reason underlying these results arises from the fact that when using the STS and DAS algorithms the sensor nodes do not report the detected targets if it is not necessary for effective navigation of actors. The information about target is transmitted from a sensor node to an actor when the distance between them is shorter. Therefore, the probability that a new target will appear closer to the selected actor before it reaches the previously reported target is diminished and there is smaller chance that the assignment of targets to actors will be non-optimal.

\section{Conclusion}

Reduction of data transmission is an important issue for the development of WSAN-based surveillance applications that require real-time data delivery, energy conservation, and effective utilization of the bandwidth-limited wireless communication medium. In this paper an approach is introduced to reduce the data transmission in WSAN by means of suppression methods that were originally intended for wireless sensor networks. Communication algorithms based on temporal, spatiotemporal, and decision aware data suppression methods are proposed for a WSAN system in which mobile actors have to capture distributed targets in the shortest possible time.

Effectiveness of the proposed data communication algorithms was verified in computational experiments by using a WSAN model. The experimental results show that the spatiotemporal and decision aware suppression methods reduce the number of data transfers and hop counts in comparison with the temporal suppression, which ensures that the actors receive complete information about targets detected in their communication ranges. Further research will be conducted to test the proposed approach in more complex network scenarios. Moreover, an interesting topic for future works is to investigate the impact of transmission failures on performance of the presented algorithms.

\end{document}